\pdfoutput=1

\documentclass[11pt]{article}

\usepackage[preprint]{acl}
\usepackage{times}
\usepackage{latexsym}
\usepackage[T1]{fontenc}
\usepackage[utf8]{inputenc}
\usepackage{microtype}
\usepackage{inconsolata}
\usepackage{graphicx}
\usepackage{multicol}
\usepackage{multirow}
\usepackage{booktabs}
\usepackage{amssymb}
\usepackage{bm}
\usepackage{pifont}
\usepackage{xcolor}
\usepackage[frozencache,cachedir=.]{minted}

\title{LLM4Ranking: An Easy-to-use Framework of Utilizing Large Language Models for Document Reranking}

\author{
 \textbf{Qi Liu\textsuperscript{1}}, 
 \textbf{Haozhe Duan\textsuperscript{1}},   
 \textbf{Yiqun Chen\textsuperscript{1}}, 
 \textbf{Quanfeng Lu\textsuperscript{2}}, \\
 \textbf{Weiwei Sun\textsuperscript{3}}, 
 \textbf{Jiaxin Mao\textsuperscript{1}}
\\
 \textsuperscript{1}Renmin University of China,
 \textsuperscript{2}Shanghai Jiao Tong University, \\
 \textsuperscript{3}Carnegie Mellon University
\\
   \texttt{qiliu6777@gmail.com, maojiaxin@gmail.com}
}

\begin{document}

\maketitle

\begin{abstract}
Utilizing large language models (LLMs) for document reranking has been a popular and promising research direction in recent years, many studies are dedicated to improving the performance and efficiency of using LLMs for reranking. Besides, it can also be applied in many real-world applications, such as search engines or retrieval-augmented generation.
In response to the growing demand for research and application in practice, we introduce a unified framework, \textbf{LLM4Ranking}, which enables users to adopt different ranking methods using open-source or closed-source API-based LLMs. Our framework provides a simple and extensible interface for document reranking with LLMs, as well as easy-to-use evaluation and fine-tuning scripts for this task. We conducted experiments based on this framework and evaluated various models and methods on several widely used datasets, providing reproducibility results on utilizing LLMs for document reranking.
Our code is publicly available at \url{https://github.com/liuqi6777/llm4ranking}.
\end{abstract}

\section{Introduction}

Document reranking is a crucial step in modern information retrieval (IR) systems. After retrieving a set of candidate documents from the corpus, the IR system will utilize a more sophisticated ranking model to re-rank these candidate documents according to their relevance to the issued query. Efficient and effective document reranking has become an important research direction in the past decades, with significant progress demonstrated in ranking models, such as learning-to-rank approaches~\cite{burges2005learning,cao2007learning,liu2009learning} and neural reranking models based on pre-trained language models~\cite{nogueira2020Passage,nogueira2019Document}. 

\begin{listing}[t]
\begin{minted}[fontsize=\small, breaklines]{python}
from llm4ranking import Reranker

reranker = Reranker(
    reranking_approach="rankgpt",
    model_type="openai", model_name="gpt-4o"
)
reranker.rerank(
    query: "query text",
    candidates: ["doc0", "doc1", "doc2", ...],
)

>> ["doc2", "doc0", "doc1", ...]
\end{minted}
\caption{Minimal usage example of LLM4Ranking. Users can leverage different reranking approaches or LLMs to rerank documents in just a few lines of code.}
\label{lst:minimal}
\end{listing}

The recent emergence of large language models (LLMs), such as GPT-4~\cite{openai2023GPT4}, PaLM~\cite{anil2023PaLM}, and Llama~\cite{touvron2023LLaMA}, has reshaped the landscape of reranking. 
With their vast pre-trained knowledge and strong reasoning abilities, LLMs offer unprecedented capabilities to capture nuanced language patterns and contextual relevance between queries and documents, and have been widely explored in reranking~\cite{zhu2023Large}. Several LLM-based reranking methods, such as RankGPT~\cite{sun2023ChatGPT}, have been proposed and proven to outperform traditional neural ranking models~\cite{sun2023ChatGPT, qin2023Large, chen2024TourRank}.
Moreover, they enable zero-shot or few-shot reranking, where models can perform well without extensive domain-specific fine-tuning.

\begin{table*}[th]
\centering
\footnotesize
\begin{tabular}{@{}l|cccccccc@{}}
\toprule
\multicolumn{1}{c|}{\multirow{2}{*}{\textbf{Framework}}}    & \multicolumn{4}{c}{\textbf{Supported Paradigms}}                                     & \multicolumn{2}{c}{\textbf{Supported LLMs}} & \textbf{Training}  & \textbf{Evaluation} \\
\multicolumn{1}{c|}{}                                       & \textit{point.}    & \textit{pair.}     & \textit{list.}     & \textit{customized} & \textit{Open}       & \textit{Closed}       &                    &                     \\ \midrule
\textbf{rank\_llm~\cite{pradeep2023RankVicuna}}             & \textbf{\ding{51}} &                    & \textbf{\ding{51}} & Hard                & Specified           & OpenAI                & \textbf{\ding{51}} & \textbf{\ding{51}}  \\
\textbf{rerankers~\cite{clavie2024rerankers}}               & \textbf{\ding{51}} &                    & \textbf{\ding{51}} & Hard                & Specified           & OpenAI                &                    &                     \\
\textbf{PyTerrier-GenRank~\cite{dhole2024PyTerrierGenRank}} &                    &                    & \textbf{\ding{51}} & Hard                & Any                 & OpenAI                &                    &                     \\
\textbf{LLM4Ranking}                                        & \textbf{\ding{51}} & \textbf{\ding{51}} & \textbf{\ding{51}} & Easy                & Any                 & OpenAI                & \textbf{\ding{51}} & \textbf{\ding{51}}  \\ \bottomrule
\end{tabular}
\caption{\textbf{Comparison between different frameworks on features.} \textit{point.} means pointwise ranking methods, and so forth. In the column of supported closed LLMs, we use OpenAI to denote the basic implementation, however, it should be noted that most LLM with APIs are compatible.}
\label{tab:comparison}
\end{table*}

While utilizing LLMs for reranking has been a promising direction for both research and real-world applications, a unified, extensible framework for experimenting with different LLM-based reranking methods and different LLMs is lacking. 
Existing frameworks have been limited in their scope, supporting only a narrow range of reranking methods or LLMs, as shown in Table~\ref{tab:comparison}. This limitation highlights the need for a flexible and comprehensive framework that can accommodate diverse combinations of methods and fine-tuning approaches to facilitate full explorations of possibilities in different areas.

To bridge this gap, we introduce LLM4Ranking, a unified framework designed to facilitate easy and systematic exploration of LLMs for document reranking. Listing~\ref{lst:minimal} shows a minimal usage example of using our framework to rerank documents. The key features include:




\paragraph{Unified and extensible interface} Users can seamlessly integrate various LLMs into their ranking pipeline with minimal effort. Such a unified interface also facilitates experimentation with different ranking strategies.

\paragraph{Support for a wide range of reranking methods} The framework integrates different popular re-ranking methods proposed recently. The framework accommodates both widely available open-source LLMs and commercial APIs, making it accessible to a broad range of users. In addition, it provides ready-to-use training codes for users to train a supervised and customized model.

\paragraph{Reproducibility and benchmarking} By providing standardized evaluation code and datasets, LLM4Ranking ensures that researchers can easily make reproducible experiments and conduct evaluations for new methods, allowing for fair comparisons across models and methodologies.

To demonstrate the capabilities and effectiveness of LLM4Ranking, we use it to evaluate both zero-shot or supervised reranking methods on multiple widely used datasets. By sharing the reproducible results, we also hope to empower researchers and practitioners to explore and advance the field of LLM-based reranking further.

In summary, our contributions are as follows:
\begin{itemize}
    \item We develop LLM4Ranking, which simplifies the integration and evaluation of LLM-based reranking methods.
    \item We evaluate the framework by performing training and evaluation experiments based on it and show its capabilities.
\end{itemize}

\section{Background and Related Work}

Large language models have demonstrated impressive effectiveness on document reranking tasks. In general, there are three main paradigms for prompting large language models: \emph{pointwise}, \emph{pairwise}, and \emph{listwise}. The pointwise approach evaluates the relevance score on one query-passage pair at a time~\cite{liang2022Holistic,sachan2023Questions, liu2024DemoRank}. The pairwise approach prompts LLM with a pair of passages to a given query to indicate which is more relevant and use aggregation methods to derive the final ranking~\cite{pradeep2021ExpandoMonoDuo,qin2023Large}. The listwise approach aims to receive a query along with a list of candidates and directly generate a ranking list based on their relevance to the query~\cite{ma2023ZeroShot,sun2023ChatGPT, liu2024Sliding}. Lots of work aimed to improve the ranking performance under these paradigms.

\begin{figure*}[t]
    \centering
    \includegraphics[width=\textwidth]{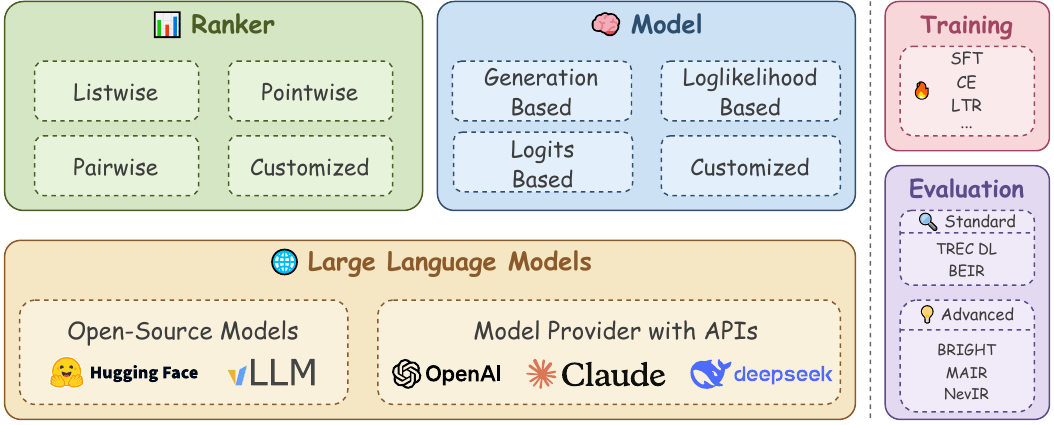}
    \caption{The overall framework of LLM4Ranking. The left part shows three core components: the backend of large language models, the ranker that holds the abstract ranking algorithm, and the specific model that used in the ranker. The right part shows the integrated features of the framework, including training and evaluation.}
    \label{fig:framework}
\end{figure*}

Beyond ranking effectiveness, research has also explored efficiency improvements, including distilling smaller models~\cite{pradeep2023RankVicuna,pradeep2023RankZephyr,zhang2023RankwithoutGPT}, passage compression~\cite{liu2024Leveraging}, or different approaches to obtain relevance score~\cite{reddy2024FIRST, chen2024Attention}. Besides, some works proposed different ranking paradigms, such as Setwise~\cite{zhuang2023Setwise} and TourRank~\cite{chen2024TourRank}, to achieve the balance of effectiveness and efficiency.

In response to the growing demand for research, it is necessary to develop a unified, extendable, and easy-to-use framework. However, as listed in Table~\ref{tab:comparison}, existing frameworks have been limited in their scope. For example, rank\_llm~\cite{pradeep2023RankVicuna} and PyTerrier-GenRank~\cite{dhole2024PyTerrierGenRank} most focused on listwise reranking, while rerankers~\cite{clavie2024rerankers} is a general framework and LLM for reranking is not its main feature. In addition, it's difficult to customize the reranking paradigms or train and evaluate the reranking models with these existing frameworks. In contrast, LLM4Ranking aims to address these issues by using a more flexible implementation, accommodating diverse LLM-based reranking methods, and supporting various training and evaluation settings, making it highly versatile and broadly applicable.

\section{The LLM4Ranking Framework}

In this section, we first present an overview of the LLM4Ranking in Section~\ref{sec:overview}, then detail the training and evaluation feature in Section~\ref{sec:training} and~\ref{sec:evaluation}.

\subsection{Overview}\label{sec:overview}

To achieve flexibility and comprehensiveness, the LLM4Ranking framework is designed as a modular system to simplify LLM-based document reranking. Basically, as Figure~\ref{fig:framework} shows, its architecture consists of three core components: \textit{LLM Interface}, \textit{Ranking Logic Abstraction}, and \textit{Model}. We present more examples in Appendix~\ref{app:example} to show how LLM, ranking logit and model can be easily combined and extended based on this framework.

\paragraph{LLM Interface} 

In LLM4Ranking, we integrate access to both open-source and proprietary LLMs to keep pace with the swift advancements. 

For open-source LLMs, we implement rich features based on the HuggingFace Transformers Library~\cite{wolf-etal-2020-transformers}, and users can load any chat-based LLMs supported in HuggingFace Transformers.\footnote{\url{https://github.com/huggingface/transformers}} In addition, we include quantization deployment strategies to enhance memory efficiency during inference, specifically bitsandbytes~\cite{dettmers20228bit} and GPTQ~\cite{frantar2023GPTQ}. Both methods facilitate 8-bit and 4-bit quantization and GPTQ additionally supports 3-bit quantization. We are also compatible with using vLLM framework~\cite{kwon2023efficient} to accelerate inference further.\footnote{\url{https://github.com/vllm-project/vllm}}

As for LLM providers with APIs, we implement the interface using the OpenAI SDK for Python and support different chat models.\footnote{\url{https://platform.openai.com/docs/overview}} Since most LLMs' API on the market are compatible with the OpenAI SDK, such as Anthropic Claude\footnote{\url{https://www.anthropic.com/}} and DeepSeek\footnote{\url{https://www.deepseek.com/}}, users can also use these LLMs in our framework.

We implement several unified interfaces for calling LLM in subsequent different ranking models, including \texttt{generate} for normal generation, \texttt{loglikelihood} to get the loglikelihood of a given target text, and \texttt{logits} to get the output logits of the specific token(s) at the last position.

\paragraph{Ranking Logic Abstraction}

In our framework, an important design principle is decoupling abstract ranking logic or paradigm (e.g., pointwise) from concrete ranking models (e.g., relevance generation). In contrast to other existing tightly coupled frameworks, this design offers the advantage that users or researchers can easily implement and evaluate new customized ranking methods.

We cover several basic ranking paradigms within LLM4Ranking to provide the most widely applicable choices, including \textit{pointwise}, \textit{pairwise}, and \textit{listwise}. Here we only implement the abstract ranking logic required by different paradigms. For example, the simplified pointwise reranker code could be:
\begin{minted}[fontsize=\small, breaklines]{python}
class PointwiseReranker:
  def rerank(
    self, query: str, candidates: list[str],
    ranking_func: Callable[[str, str], float],
  ):
    return sorted(candidates, key=lambda doc: ranking_func(query, doc), reverse=True)
\end{minted}
Then we can pass different reranking models (which will be elaborated in the following section) to the \texttt{rerank} function through the \texttt{ranking\_func} argument, without needing to concern how the LLM derives this score here.

Additionally, users can easily implement other ranking logic following a similar template. As an example, we also include TourRank~\cite{chen2024TourRank}, a selection paradigm inspired by the Tournament mechanism, in our framework.

\paragraph{Model}

The \textit{Model} component in LLM4Ranking provides the concrete implementation of different ranking models. Corresponding to the three interfaces implemented in the LLM module, we categorize models into three primary approaches based on how they interact with LLMs:
\begin{itemize}
\item \textit{Generation-based Model.} 
This approach formulates document ranking as a text generation task, where the LLM generates a relevance score, justification, or ranking order based on the given query and candidate documents. Methods such as RankGPT~\cite{sun2023ChatGPT} and TourRank~\cite{chen2024TourRank} fall under this category, as they rely on LLMs' inherent ability to generate structured ranking responses.

\item \textit{Log-likelihood-based Model.} 
Instead of generating free-form text, this approach computes the ranking score by measuring the log-likelihood of a specific target text. This method is useful for evaluating how confidently an LLM assigns relevance to a document, and it enables scoring mechanisms such as query generation~\cite{sachan2023Questions} and fine-grained relevance generation~\cite{zhuang2023Yes}.

\item \textit{Logits-based Model.} 
This approach directly utilizes the LLM's output logits at the last token position to assess relevance signals. By extracting the probability distributions over specific tokens, models leveraging this method can perform ranking decisions. For example, relevance generation~\cite{liang2022Holistic} takes the logit of ``yes'' as the relevance score, PRP~\cite{qin2023Large} takes the document with a higher logit of the identifier as the more relevant document, and FIRST~\cite{reddy2024FIRST} directly ranks a list of documents according to the logits of the identifiers.
\end{itemize}
These three model types collectively offer a flexible and extensible foundation for document reranking, and the above implemented models allow users to experiment with different methodologies depending on their specific needs and computational constraints. 

Beyond these predefined models, LLM4Ranking enables easy implementation of customized new ranking models through its modular design and unified LLM interface. With structured templates and utilities, the framework simplifies development, allowing researchers to prototype and evaluate ranking approaches without managing low-level LLM interactions.

\subsection{Training}\label{sec:training}

In addition to the core components, LLM4Ranking provides a set of tools to drive the training of different ranking models. Specifically, for different type of models, there are two different training programs. Firstly, for generation-based and log-likelihood-based models, we offer out-of-the-box scripts for the standard Supervised Fine-tuning (SFT) pipeline, which can be directly used in the command line. For example, one can train a listwise reranker using the following commands:
\begin{minted}[fontsize=\small, breaklines]{bash}
torchrun llm4ranking/training/sft/train.py \
    --model_name_or_path Qwen/Qwen2.5-0.5B-Instruct \
    --data_path /path/to/your_sft_dataset.jsonl \
    --output_dir /path/to/your_model_name
\end{minted}
The other training arguments are the same as those of \texttt{TrainingArguments} in Huggingface Transformers. The SFT dataset should be in the format of conversations:
\begin{minted}[fontsize=\small, breaklines]{json}
{
    "id": "<data sample id>",
    "messages": [
        {"role": "system", "content": "<system message>"},
        {"role": "user", "content": "<user message>"},
        {"role": "assistant", "content": "<assistant message>"},
        ...
    ]
}
\end{minted}
We processed the data generated from RankGPT provided by~\citet{pradeep2023RankZephyr} for distilling the smaller listwise model, and users can construct custom training datasets in the above format. Additionally, PEFT such as Lora~\cite{hu2021LoRA} is also supported.

Secondly, for logits-based models, such as Relevance Generation, the training process is entirely different from SFT. Therefore, we refer to the cross-encode and implement another set of training codes for these models. Specifically, we implement a new \texttt{Trainer} and a set of loss functions, including widely used Cross-Entropy loss, Margin-MSE loss and learning-to-rank (LTR) losses such as RankNet~\cite{burges2005learning}. A training example is shown as follows:
\begin{minted}[fontsize=\small, breaklines]{bash}
torchrun llm4ranking/training/logits/train.py \
    --model_name_or_path Qwen/Qwen2.5-0.5B-Instruct \
    --data_path /path/to/your_dataset.jsonl \
    --output_dir /path/to/your_model_name \
    --loss_type cross_entropy \
    --num_negatives 7
\end{minted}
where \texttt{--loss\_type} specifies the loss function to be used for training, while \texttt{--num\_negatives} determines the number of negative examples to be used. For the data format, we refer to the settings of Tevatron~\cite{gao2022Tevatron} and recommend that users directly use the processed datasets published by them.

\subsection{Evaluation}\label{sec:evaluation}

LLM4ranking supports a wide range of evaluation settings. We cover multiple popular academic datasets for evaluating reranker, including the standard retrieval dataset such as TREC DL~\cite{craswell2020overview} and BEIR~\cite{thakur2021BEIR}, as well as advanced datasets: MAIR~\cite{sun2024MAIR} for instruction-following retrieval, NevIR for negation retrieval~\cite{weller2024NevIR}, and Bright~\cite{su2024BRIGHT} for reasoning-intensive retrieval. For each dataset, we performed standard operations. Specifically, we followed the commonly used settings and used BM25 as the retrieval model to retrieve the top 100 candidate documents. We also publicly released a unified format for users to evaluate the ranking model in an easy and unified manner.

We support conducting evaluation experiments through the command line:
\begin{minted}[fontsize=\small, breaklines]{bash}
python -m llm4ranking.evaluation.evaluator \
    --model_type openai \
    --model_args model=gpt-4o,api_key=sk-xxxx \
    --model_fw_args temperature=0 \
    --reranking_approach rankgpt \
    --reranking_args window_size=20,step=10 \
    --datasets dl19 dl20 \
    --retriever bm25 \
    --topk 100 \
    --output_dir path/to/your/folder
\end{minted}
where \texttt{--model\_type} and \texttt{--model\_args} decide the LLM to evaluate, \texttt{--reranking\_approach} and \texttt{--reranking\_args} decide the reranking model. We also provide a wrapped function interface for evaluation, with the same arguments as the command line usage. 

The results will be saved under the specified path, including a text file that stores the ranking output in TREC format, and a JSON file that stores the evaluation metrics (MAP, NDCG, and Recall) and detailed running records. The records include the reranking latency, the number of processed and generated tokens, and the output of the LLM, and could be used for further analysis.

\section{Experiments}

\subsection{Experimental Setup}

To demonstrate LLM4Ranking's capability, we conduct experiments based on the framework. 
Firstly, we evaluate several baselines in zero-shot manner, including pointwise method \textit{Relevance generation}~\cite{liang2022Holistic}, pairwise method \textit{PRP-Heapsort}~\cite{qin2023Large}, listwise method \textit{RankGPT}~\cite{sun2023ChatGPT}, and selection-based method \textit{TourRank-1}~\cite{chen2024TourRank}. 
We use open-source instruct models (Llama 3.1 series models~\cite{grattafiori2024llama3} and Qwen 2.5 series models~\cite{qwen2024Qwen25}) and proprietary models with APIs (OpenAI GPT-4o and GPT-4o-mini) to perform the above methods. 

Secondly, we train and evaluate supervised pointwise and listwise models based on Qwen 2.5 series models but with smaller sizes ranging from 0.5B to 7B. We fine-tuned pointwise rerankers for using MS MARCO training set. For listwise rerankers, following~\citet{pradeep2023RankZephyr}, we distill from RankGPT-4~\cite{sun2023ChatGPT}. Note that we are only showcasing the training feature of the framework here, and the hyperparameter tuning and data engineering are beyond the scope of this paper. These fine-tuned models are also open-sourced.

For all experiments, we use the test sets of TREC DL benchmarks~\cite{craswell2020overview}. Following~\citet{sun2023ChatGPT}, we rerank the top 100 candidates obtained from BM25 and use nDCG@10 as the metric to evaluate the reranking results. More details can be found in Appendix~\ref{sec:appendix-exp}.

\subsection{Results}

\paragraph{Zero-shot Evaluation Results}

\begin{table}[t]
\small
\centering
\begin{tabular}{@{}llcc@{}}
\toprule
\multicolumn{1}{l|}{\textbf{LLM}}                       & \multicolumn{1}{l|}{\textbf{Method}} & \textbf{DL19}   & \textbf{DL20}        \\ \midrule
\multicolumn{1}{l|}{-}                                  & \multicolumn{1}{l|}{BM25}            & 0.5058               & 0.4796               \\ \midrule
\multicolumn{4}{c}{\textit{Open-Source LLMs}}                                                                                                \\ \midrule
\multicolumn{1}{l|}{\multirow{4}{*}{Llama-3.1-8B}}      & \multicolumn{1}{l|}{RelGen}          & 0.6548               & 0.6023               \\
\multicolumn{1}{l|}{}                                   & \multicolumn{1}{l|}{PRP-Heap}        & 0.6086               & 0.5465               \\
\multicolumn{1}{l|}{}                                   & \multicolumn{1}{l|}{RankGPT}         & 0.6775               & 0.6529               \\
\multicolumn{1}{l|}{}                                   & \multicolumn{1}{l|}{TourRank-1}      & 0.6721               & 0.6314               \\ \midrule
\multicolumn{1}{l|}{\multirow{4}{*}{Qwen-2.5-7B}}       & \multicolumn{1}{l|}{RelGen}          & 0.5239               & 0.5243               \\
\multicolumn{1}{l|}{}                                   & \multicolumn{1}{l|}{PRP-Heap}        & 0.7073               & 0.6597               \\
\multicolumn{1}{l|}{}                                   & \multicolumn{1}{l|}{RankGPT}         & 0.6870               & 0.6386               \\
\multicolumn{1}{l|}{}                                   & \multicolumn{1}{l|}{TourRank-1}      & 0.6704               & 0.6051               \\ \midrule
\multicolumn{4}{c}{\textit{LLM Provider with APIs}}                                                                                          \\ \midrule
\multicolumn{1}{l|}{\multirow{2}{*}{GPT-4o}}            & \multicolumn{1}{l|}{RankGPT}         & 0.7506               & \textbf{0.7106}      \\
\multicolumn{1}{l|}{}                                   & \multicolumn{1}{l|}{TourRank-1}      & 0.7289               & 0.6712               \\ \midrule
\multicolumn{1}{l|}{\multirow{2}{*}{Claude-3.7-Sonnet}} & \multicolumn{1}{l|}{RankGPT}         & 0.7319               & 0.7009               \\
\multicolumn{1}{l|}{}                                   & \multicolumn{1}{l|}{TourRank-1}      & 0.7303               & 0.6677               \\ \midrule
\multicolumn{1}{l|}{\multirow{2}{*}{DeepSeek-V3}}       & \multicolumn{1}{l|}{RankGPT}         & \textbf{0.7590}      & 0.7064               \\
\multicolumn{1}{l|}{}                                   & \multicolumn{1}{l|}{TourRank-1}      & 0.7176               & 0.6854               \\ \bottomrule
\end{tabular}
\caption{The zero-shot results of different reranking methods with different LLMs using LLM4Ranking.}
\label{tab:zero-shot}
\end{table}

Table~\ref{tab:zero-shot} presents the zero-shot reranking performance of various LLM-based methods on the TREC DL19 and DL20 benchmarks. Among both open-source models and LLMs accessed via APIs, RankGPT a high effectiveness across both datasets, notably achieving 0.7506 nDCG@10 on DL19 and 0.7106 on DL20 using GPT-4o. Although we only perform 1 tournament, Tourrank-1's performance follows closely behind RankGPT. For the two methods of RelGen and PRP-Heap, different LLMs show different performances. The results of Qwen-2.5-7B using PRP-Heap even surpasses RankGPT, but perform poorly on RelGen; however, LLama-3-8B is exactly the opposite.

As for the comparison between models, API-based models generally surpass their open-source counterparts with smaller parameter sizes, suggesting that more advanced and larger-scale proprietary LLMs provide superior reranking performance.

\paragraph{Supervised Evaluation Results}

Table~\ref{tab:training-result} summarizes the performance of supervised rerankers fine-tuned on the MS MARCO dataset. As expected, performance improves with increasing model size. For the pointwise RelGen approach, the NDCG@10 score steadily rises from 0.7139 (Qwen-2.5-0.5B) to 0.7380 (Qwen-2.5-7B) on DL19, while achieving 0.6551 to 0.6768 on DL20. Similarly, for the listwise RankGPT method, the Qwen-2.5-7B model outperforms its smaller counterparts, reaching 0.7467 on DL19 and 0.6903 on DL20. In general, RelGen has a higher ranking performance when the model size is smaller, however, with the number of parameters increasing, it may be not as good as the listwise method.

Comparing zero-shot and supervised performance, we observe that fine-tuned smaller models such as Qwen-2.5-7B can achieve results comparable to or exceeding some zero-shot LLM-based rerankers. This highlights the effectiveness of task-specific fine-tuning, particularly when computational constraints limit the deployment of larger proprietary models.

\begin{table}[t]
\centering
\small
\begin{tabular}{@{}ll|cc@{}}
\toprule
                                  & \textbf{LLM}   & \textbf{DL 19} & \textbf{DL 20} \\
\midrule
\multirow{4}{*}{\textbf{RelGen}}  & Qwen-2.5-0.5B  & 0.7139        & 0.6551        \\
                                  & Qwen-2.5-1.5B  & 0.7295        & 0.6875        \\
                                  & Qwen-2.5-3B    & 0.7353        & \textbf{0.6962}        \\
                                  & Qwen-2.5-7B    & \textbf{0.7380}        & 0.6768        \\
\midrule
\multirow{4}{*}{\textbf{RankGPT}} & Qwen-2.5-0.5B  & 0.6220        & 0.5832        \\
                                  & Qwen-2.5-1.5B  & 0.7266        & 0.6748        \\
                                  & Qwen-2.5-3B    & 0.7352        & 0.6890        \\
                                  & Qwen-2.5-7B    & \textbf{0.7467}        & \textbf{0.6903}        \\
\bottomrule
\end{tabular}
\caption{The results of supervised models.}
\label{tab:training-result}
\end{table}


\section{Conclusion}

In this paper, we present LLM4Ranking, an easy-to-use toolkit for leveraging LLMs for document reranking, which provides support for various reranking methods and LLMs, benchmark evaluation, and training strategies within a unified, simple, and flexible framework.
We demonstrate LLM4Ranking's capabilities by constructing massive experiments, illustrating its effectiveness in training and evaluation.
We believe that LLM4Ranking will serve as a useful toolkit for both academics and the community in evaluating LLM-based rerankers or real-world applications such as RAG, thereby contributing to the advancement of natural language processing and information retrieval fields.


\bibliography{reference, addition}

\begin{thebibliography}{41}
\providecommand{\natexlab}[1]{#1}

\bibitem[{Anil et~al.(2023)Anil, Dai, Firat, Johnson, Lepikhin, Passos, Shakeri, Taropa, Bailey, Chen, Chu, Clark, Shafey, Huang, {Meier-Hellstern}, Mishra, Moreira, Omernick, Robinson, Ruder, Tay, Xiao, Xu, Zhang, Abrego, Ahn, Austin, Barham, Botha, Bradbury, Brahma, Brooks, Catasta, Cheng, Cherry, {Choquette-Choo}, Chowdhery, Crepy, Dave, Dehghani, Dev, Devlin, D{\'i}az, Du, Dyer, Feinberg, Feng, Fienber, Freitag, Garcia, Gehrmann, Gonzalez, {Gur-Ari}, Hand, Hashemi, Hou, Howland, Hu, Hui, Hurwitz, Isard, Ittycheriah, Jagielski, Jia, Kenealy, Krikun, Kudugunta, Lan, Lee, Lee, Li, Li, Li, Li, Li, Lim, Lin, Liu, Liu, Maggioni, Mahendru, Maynez, Misra, Moussalem, Nado, Nham, Ni, Nystrom, Parrish, Pellat, Polacek, Polozov, Pope, Qiao, Reif, Richter, Riley, Ros, Roy, Saeta, Samuel, Shelby, Slone, Smilkov, So, Sohn, Tokumine, Valter, Vasudevan, Vodrahalli, Wang, Wang, Wang, Wang, Wieting, Wu, Xu, Xu, Xue, Yin, Yu, Zhang, Zheng, Zheng, Zhou, Zhou, Petrov, and Wu}]{anil2023PaLM}
Rohan Anil, Andrew~M. Dai, Orhan Firat, Melvin Johnson, Dmitry Lepikhin, Alexandre Passos, Siamak Shakeri, Emanuel Taropa, Paige Bailey, Zhifeng Chen, Eric Chu, Jonathan~H. Clark, Laurent~El Shafey, Yanping Huang, Kathy {Meier-Hellstern}, Gaurav Mishra, Erica Moreira, Mark Omernick, Kevin Robinson, Sebastian Ruder, Yi~Tay, Kefan Xiao, Yuanzhong Xu, Yujing Zhang, Gustavo~Hernandez Abrego, Junwhan Ahn, Jacob Austin, Paul Barham, Jan Botha, James Bradbury, Siddhartha Brahma, Kevin Brooks, Michele Catasta, Yong Cheng, Colin Cherry, Christopher~A. {Choquette-Choo}, Aakanksha Chowdhery, Cl{\'e}ment Crepy, Shachi Dave, Mostafa Dehghani, Sunipa Dev, Jacob Devlin, Mark D{\'i}az, Nan Du, Ethan Dyer, Vlad Feinberg, Fangxiaoyu Feng, Vlad Fienber, Markus Freitag, Xavier Garcia, Sebastian Gehrmann, Lucas Gonzalez, Guy {Gur-Ari}, Steven Hand, Hadi Hashemi, Le~Hou, Joshua Howland, Andrea Hu, Jeffrey Hui, Jeremy Hurwitz, Michael Isard, Abe Ittycheriah, Matthew Jagielski, Wenhao Jia, Kathleen Kenealy, Maxim Krikun, Sneha
  Kudugunta, Chang Lan, Katherine Lee, Benjamin Lee, Eric Li, Music Li, Wei Li, YaGuang Li, Jian Li, Hyeontaek Lim, Hanzhao Lin, Zhongtao Liu, Frederick Liu, Marcello Maggioni, Aroma Mahendru, Joshua Maynez, Vedant Misra, Maysam Moussalem, Zachary Nado, John Nham, Eric Ni, Andrew Nystrom, Alicia Parrish, Marie Pellat, Martin Polacek, Alex Polozov, Reiner Pope, Siyuan Qiao, Emily Reif, Bryan Richter, Parker Riley, Alex~Castro Ros, Aurko Roy, Brennan Saeta, Rajkumar Samuel, Renee Shelby, Ambrose Slone, Daniel Smilkov, David~R. So, Daniel Sohn, Simon Tokumine, Dasha Valter, Vijay Vasudevan, Kiran Vodrahalli, Xuezhi Wang, Pidong Wang, Zirui Wang, Tao Wang, John Wieting, Yuhuai Wu, Kelvin Xu, Yunhan Xu, Linting Xue, Pengcheng Yin, Jiahui Yu, Qiao Zhang, Steven Zheng, Ce~Zheng, Weikang Zhou, Denny Zhou, Slav Petrov, and Yonghui Wu. 2023.
\newblock \href {https://arxiv.org/abs/2305.10403} {{{PaLM}} 2 {{Technical Report}}}.
\newblock \emph{Preprint}, arXiv:2305.10403.

\bibitem[{Burges et~al.(2005)Burges, Shaked, Renshaw, Lazier, Deeds, Hamilton, and Hullender}]{burges2005learning}
Chris Burges, Tal Shaked, Erin Renshaw, Ari Lazier, Matt Deeds, Nicole Hamilton, and Greg Hullender. 2005.
\newblock Learning to rank using gradient descent.
\newblock In \emph{Proceedings of the 22nd international conference on Machine learning}, pages 89--96.

\bibitem[{Cao et~al.(2007)Cao, Qin, Liu, Tsai, and Li}]{cao2007learning}
Zhe Cao, Tao Qin, Tie-Yan Liu, Ming-Feng Tsai, and Hang Li. 2007.
\newblock Learning to rank: from pairwise approach to listwise approach.
\newblock In \emph{Proceedings of the 24th international conference on Machine learning}, pages 129--136.

\bibitem[{Chen et~al.(2024{\natexlab{a}})Chen, Guti{\'e}rrez, and Su}]{chen2024Attention}
Shijie Chen, Bernal~Jim{\'e}nez Guti{\'e}rrez, and Yu~Su. 2024{\natexlab{a}}.
\newblock \href {https://arxiv.org/abs/2410.02642} {Attention in {{Large Language Models Yields Efficient Zero-Shot Re-Rankers}}}.
\newblock \emph{Preprint}, arXiv:2410.02642.

\bibitem[{Chen et~al.(2024{\natexlab{b}})Chen, Liu, Zhang, Sun, Shi, Mao, and Yin}]{chen2024TourRank}
Yiqun Chen, Qi~Liu, Yi~Zhang, Weiwei Sun, Daiting Shi, Jiaxin Mao, and Dawei Yin. 2024{\natexlab{b}}.
\newblock \href {https://arxiv.org/abs/2406.11678} {{{TourRank}}: {{Utilizing Large Language Models}} for {{Documents Ranking}} with a {{Tournament-Inspired Strategy}}}.
\newblock \emph{Preprint}, arXiv:2406.11678.

\bibitem[{Clavi{\'e}(2024)}]{clavie2024rerankers}
Benjamin Clavi{\'e}. 2024.
\newblock \href {https://doi.org/10.48550/arXiv.2408.17344} {Rerankers: {{A Lightweight Python Library}} to {{Unify Ranking Methods}}}.
\newblock \emph{Preprint}, arXiv:2408.17344.

\bibitem[{Craswell et~al.(2020)Craswell, Mitra, Yilmaz, Campos, and Voorhees}]{craswell2020overview}
Nick Craswell, Bhaskar Mitra, Emine Yilmaz, Daniel Campos, and Ellen~M Voorhees. 2020.
\newblock Overview of the trec 2019 deep learning track.
\newblock \emph{arXiv preprint arXiv:2003.07820}.

\bibitem[{Dettmers et~al.(2022)Dettmers, Lewis, Shleifer, and Zettlemoyer}]{dettmers20228bit}
Tim Dettmers, Mike Lewis, Sam Shleifer, and Luke Zettlemoyer. 2022.
\newblock \href {https://doi.org/10.48550/arXiv.2110.02861} {8-bit {{Optimizers}} via {{Block-wise Quantization}}}.
\newblock \emph{Preprint}, arXiv:2110.02861.

\bibitem[{Dhole(2024)}]{dhole2024PyTerrierGenRank}
Kaustubh~D. Dhole. 2024.
\newblock \href {https://doi.org/10.48550/arXiv.2412.05339} {{{PyTerrier-GenRank}}: {{The PyTerrier Plugin}} for {{Reranking}} with {{Large Language Models}}}.
\newblock \emph{Preprint}, arXiv:2412.05339.

\bibitem[{Frantar et~al.(2023)Frantar, Ashkboos, Hoefler, and Alistarh}]{frantar2023GPTQ}
Elias Frantar, Saleh Ashkboos, Torsten Hoefler, and Dan Alistarh. 2023.
\newblock \href {https://doi.org/10.48550/arXiv.2210.17323} {{{GPTQ}}: {{Accurate Post-Training Quantization}} for {{Generative Pre-trained Transformers}}}.
\newblock \emph{Preprint}, arXiv:2210.17323.

\bibitem[{Gao et~al.(2022)Gao, Ma, Lin, and Callan}]{gao2022Tevatron}
Luyu Gao, Xueguang Ma, Jimmy Lin, and Jamie Callan. 2022.
\newblock \href {https://doi.org/10.48550/arXiv.2203.05765} {Tevatron: {{An Efficient}} and {{Flexible Toolkit}} for {{Dense Retrieval}}}.
\newblock \emph{Preprint}, arXiv:2203.05765.

\bibitem[{Grattafiori et~al.(2024)Grattafiori, Dubey, Jauhri, Pandey, Kadian, Al-Dahle, Letman, Mathur, Schelten, Vaughan et~al.}]{grattafiori2024llama3}
Aaron Grattafiori, Abhimanyu Dubey, Abhinav Jauhri, Abhinav Pandey, Abhishek Kadian, Ahmad Al-Dahle, Aiesha Letman, Akhil Mathur, Alan Schelten, Alex Vaughan, et~al. 2024.
\newblock The llama 3 herd of models.
\newblock \emph{arXiv preprint arXiv:2407.21783}.

\bibitem[{Hu et~al.(2021)Hu, Shen, Wallis, {Allen-Zhu}, Li, Wang, Wang, and Chen}]{hu2021LoRA}
Edward~J. Hu, Yelong Shen, Phillip Wallis, Zeyuan {Allen-Zhu}, Yuanzhi Li, Shean Wang, Lu~Wang, and Weizhu Chen. 2021.
\newblock \href {https://doi.org/10.48550/arXiv.2106.09685} {{{LoRA}}: {{Low-Rank Adaptation}} of {{Large Language Models}}}.
\newblock \emph{Preprint}, arXiv:2106.09685.

\bibitem[{Kwon et~al.(2023)Kwon, Li, Zhuang, Sheng, Zheng, Yu, Gonzalez, Zhang, and Stoica}]{kwon2023efficient}
Woosuk Kwon, Zhuohan Li, Siyuan Zhuang, Ying Sheng, Lianmin Zheng, Cody~Hao Yu, Joseph~E. Gonzalez, Hao Zhang, and Ion Stoica. 2023.
\newblock Efficient memory management for large language model serving with pagedattention.
\newblock In \emph{Proceedings of the ACM SIGOPS 29th Symposium on Operating Systems Principles}.

\bibitem[{Liang et~al.(2022)Liang, Bommasani, Lee, Tsipras, Soylu, Yasunaga, Zhang, Narayanan, Wu, Kumar, Newman, Yuan, Yan, Zhang, Cosgrove, Manning, R{\'e}, {Acosta-Navas}, Hudson, Zelikman, Durmus, Ladhak, Rong, Ren, Yao, Wang, Santhanam, Orr, Zheng, Yuksekgonul, Suzgun, Kim, Guha, Chatterji, Khattab, Henderson, Huang, Chi, Xie, Santurkar, Ganguli, Hashimoto, Icard, Zhang, Chaudhary, Wang, Li, Mai, Zhang, and Koreeda}]{liang2022Holistic}
Percy Liang, Rishi Bommasani, Tony Lee, Dimitris Tsipras, Dilara Soylu, Michihiro Yasunaga, Yian Zhang, Deepak Narayanan, Yuhuai Wu, Ananya Kumar, Benjamin Newman, Binhang Yuan, Bobby Yan, Ce~Zhang, Christian Cosgrove, Christopher~D. Manning, Christopher R{\'e}, Diana {Acosta-Navas}, Drew~A. Hudson, Eric Zelikman, Esin Durmus, Faisal Ladhak, Frieda Rong, Hongyu Ren, Huaxiu Yao, Jue Wang, Keshav Santhanam, Laurel Orr, Lucia Zheng, Mert Yuksekgonul, Mirac Suzgun, Nathan Kim, Neel Guha, Niladri Chatterji, Omar Khattab, Peter Henderson, Qian Huang, Ryan Chi, Sang~Michael Xie, Shibani Santurkar, Surya Ganguli, Tatsunori Hashimoto, Thomas Icard, Tianyi Zhang, Vishrav Chaudhary, William Wang, Xuechen Li, Yifan Mai, Yuhui Zhang, and Yuta Koreeda. 2022.
\newblock \href {https://doi.org/10.48550/arXiv.2211.09110} {Holistic {{Evaluation}} of {{Language Models}}}.
\newblock \emph{Preprint}, arXiv:2211.09110.

\bibitem[{Liu et~al.(2024{\natexlab{a}})Liu, Wang, Wang, and Mao}]{liu2024Leveraging}
Qi~Liu, Bo~Wang, Nan Wang, and Jiaxin Mao. 2024{\natexlab{a}}.
\newblock \href {https://arxiv.org/abs/2406.14848} {Leveraging {{Passage Embeddings}} for {{Efficient Listwise Reranking}} with {{Large Language Models}}}.
\newblock \emph{Preprint}, arXiv:2406.14848.

\bibitem[{Liu et~al.(2009)}]{liu2009learning}
Tie-Yan Liu et~al. 2009.
\newblock Learning to rank for information retrieval.
\newblock \emph{Foundations and Trends{\textregistered} in Information Retrieval}, 3(3):225--331.

\bibitem[{Liu et~al.(2024{\natexlab{b}})Liu, Ma, Zhu, Zhao, Wang, Yin, and Dou}]{liu2024Sliding}
Wenhan Liu, Xinyu Ma, Yutao Zhu, Ziliang Zhao, Shuaiqiang Wang, Dawei Yin, and Zhicheng Dou. 2024{\natexlab{b}}.
\newblock \href {https://doi.org/10.48550/arXiv.2412.14574} {Sliding {{Windows Are Not}} the {{End}}: {{Exploring Full Ranking}} with {{Long-Context Large Language Models}}}.
\newblock \emph{Preprint}, arXiv:2412.14574.

\bibitem[{Liu et~al.(2024{\natexlab{c}})Liu, Zhu, and Dou}]{liu2024DemoRank}
Wenhan Liu, Yutao Zhu, and Zhicheng Dou. 2024{\natexlab{c}}.
\newblock \href {https://arxiv.org/abs/2406.16332} {{{DemoRank}}: {{Selecting Effective Demonstrations}} for {{Large Language Models}} in {{Ranking Task}}}.
\newblock \emph{Preprint}, arXiv:2406.16332.

\bibitem[{Ma et~al.(2023)Ma, Zhang, Pradeep, and Lin}]{ma2023ZeroShot}
Xueguang Ma, Xinyu Zhang, Ronak Pradeep, and Jimmy Lin. 2023.
\newblock \href {https://arxiv.org/abs/2305.02156} {Zero-{{Shot Listwise Document Reranking}} with a {{Large Language Model}}}.
\newblock \emph{Preprint}, arXiv:2305.02156.

\bibitem[{Nogueira and Cho(2020)}]{nogueira2020Passage}
Rodrigo Nogueira and Kyunghyun Cho. 2020.
\newblock \href {https://doi.org/10.48550/arXiv.1901.04085} {Passage {{Re-ranking}} with {{BERT}}}.
\newblock \emph{Preprint}, arXiv:1901.04085.

\bibitem[{Nogueira et~al.(2019)Nogueira, Yang, Lin, and Cho}]{nogueira2019Document}
Rodrigo Nogueira, Wei Yang, Jimmy Lin, and Kyunghyun Cho. 2019.
\newblock \href {https://doi.org/10.48550/arXiv.1904.08375} {Document {{Expansion}} by {{Query Prediction}}}.
\newblock \emph{Preprint}, arXiv:1904.08375.

\bibitem[{OpenAI(2023)}]{openai2023GPT4}
OpenAI. 2023.
\newblock \href {https://doi.org/10.48550/arXiv.2303.08774} {{{GPT-4 Technical Report}}}.
\newblock \emph{Preprint}, arXiv:2303.08774.

\bibitem[{Pradeep et~al.(2021)Pradeep, Nogueira, and Lin}]{pradeep2021ExpandoMonoDuo}
Ronak Pradeep, Rodrigo Nogueira, and Jimmy Lin. 2021.
\newblock \href {https://doi.org/10.48550/arXiv.2101.05667} {The {{Expando-Mono-Duo Design Pattern}} for {{Text Ranking}} with {{Pretrained Sequence-to-Sequence Models}}}.
\newblock \emph{Preprint}, arXiv:2101.05667.

\bibitem[{Pradeep et~al.(2023{\natexlab{a}})Pradeep, Sharifymoghaddam, and Lin}]{pradeep2023RankVicuna}
Ronak Pradeep, Sahel Sharifymoghaddam, and Jimmy Lin. 2023{\natexlab{a}}.
\newblock \href {https://arxiv.org/abs/2309.15088} {{{RankVicuna}}: {{Zero-Shot Listwise Document Reranking}} with {{Open-Source Large Language Models}}}.
\newblock \emph{Preprint}, arXiv:2309.15088.

\bibitem[{Pradeep et~al.(2023{\natexlab{b}})Pradeep, Sharifymoghaddam, and Lin}]{pradeep2023RankZephyr}
Ronak Pradeep, Sahel Sharifymoghaddam, and Jimmy Lin. 2023{\natexlab{b}}.
\newblock \href {https://doi.org/10.48550/arXiv.2312.02724} {{{RankZephyr}}: {{Effective}} and {{Robust Zero-Shot Listwise Reranking}} is a {{Breeze}}!}
\newblock \emph{Preprint}, arXiv:2312.02724.

\bibitem[{Qin et~al.(2023)Qin, Jagerman, Hui, Zhuang, Wu, Shen, Liu, Liu, Metzler, Wang, and Bendersky}]{qin2023Large}
Zhen Qin, Rolf Jagerman, Kai Hui, Honglei Zhuang, Junru Wu, Jiaming Shen, Tianqi Liu, Jialu Liu, Donald Metzler, Xuanhui Wang, and Michael Bendersky. 2023.
\newblock \href {https://doi.org/10.48550/arXiv.2306.17563} {Large {{Language Models}} are {{Effective Text Rankers}} with {{Pairwise Ranking Prompting}}}.
\newblock \emph{Preprint}, arXiv:2306.17563.

\bibitem[{Qwen et~al.(2024)Qwen, Yang, Yang, Zhang, Hui, Zheng, Yu, Li, Liu, Huang, Wei, Lin, Yang, Tu, Zhang, Yang, Yang, Zhou, Lin, Dang, Lu, Bao, Yang, Yu, Li, Xue, Zhang, Zhu, Men, Lin, Li, Xia, Ren, Ren, Fan, Su, Zhang, Wan, Liu, Cui, Zhang, and Qiu}]{qwen2024Qwen25}
Qwen, An~Yang, Baosong Yang, Beichen Zhang, Binyuan Hui, Bo~Zheng, Bowen Yu, Chengyuan Li, Dayiheng Liu, Fei Huang, Haoran Wei, Huan Lin, Jian Yang, Jianhong Tu, Jianwei Zhang, Jianxin Yang, Jiaxi Yang, Jingren Zhou, Junyang Lin, Kai Dang, Keming Lu, Keqin Bao, Kexin Yang, Le~Yu, Mei Li, Mingfeng Xue, Pei Zhang, Qin Zhu, Rui Men, Runji Lin, Tianhao Li, Tingyu Xia, Xingzhang Ren, Xuancheng Ren, Yang Fan, Yang Su, Yichang Zhang, Yu~Wan, Yuqiong Liu, Zeyu Cui, Zhenru Zhang, and Zihan Qiu. 2024.
\newblock \href {https://doi.org/10.48550/arXiv.2412.15115} {Qwen2.5 {{Technical Report}}}.
\newblock \emph{Preprint}, arXiv:2412.15115.

\bibitem[{Reddy et~al.(2024)Reddy, Doo, Xu, Sultan, Swain, Sil, and Ji}]{reddy2024FIRST}
Revanth~Gangi Reddy, JaeHyeok Doo, Yifei Xu, Md~Arafat Sultan, Deevya Swain, Avirup Sil, and Heng Ji. 2024.
\newblock \href {https://arxiv.org/abs/2406.15657} {{{FIRST}}: {{Faster Improved Listwise Reranking}} with {{Single Token Decoding}}}.
\newblock \emph{Preprint}, arXiv:2406.15657.

\bibitem[{Sachan et~al.(2023)Sachan, Lewis, Yogatama, Zettlemoyer, Pineau, and Zaheer}]{sachan2023Questions}
Devendra~Singh Sachan, Mike Lewis, Dani Yogatama, Luke Zettlemoyer, Joelle Pineau, and Manzil Zaheer. 2023.
\newblock \href {https://doi.org/10.48550/arXiv.2206.10658} {Questions {{Are All You Need}} to {{Train}} a {{Dense Passage Retriever}}}.
\newblock \emph{Preprint}, arXiv:2206.10658.

\bibitem[{Su et~al.(2024)Su, Yen, Xia, Shi, Muennighoff, Wang, Liu, Shi, Siegel, Tang, Sun, Yoon, Arik, Chen, and Yu}]{su2024BRIGHT}
Hongjin Su, Howard Yen, Mengzhou Xia, Weijia Shi, Niklas Muennighoff, Han-yu Wang, Haisu Liu, Quan Shi, Zachary~S. Siegel, Michael Tang, Ruoxi Sun, Jinsung Yoon, Sercan~O. Arik, Danqi Chen, and Tao Yu. 2024.
\newblock \href {https://arxiv.org/abs/2407.12883} {{{BRIGHT}}: {{A Realistic}} and {{Challenging Benchmark}} for {{Reasoning-Intensive Retrieval}}}.
\newblock \emph{Preprint}, arXiv:2407.12883.

\bibitem[{Sun et~al.(2024)Sun, Shi, Wu, Yan, Ma, Liu, Cao, Yin, and Ren}]{sun2024MAIR}
Weiwei Sun, Zhengliang Shi, Jiulong Wu, Lingyong Yan, Xinyu Ma, Yiding Liu, Min Cao, Dawei Yin, and Zhaochun Ren. 2024.
\newblock \href {https://doi.org/10.48550/arXiv.2410.10127} {{{MAIR}}: {{A Massive Benchmark}} for {{Evaluating Instructed Retrieval}}}.
\newblock \emph{Preprint}, arXiv:2410.10127.

\bibitem[{Sun et~al.(2023)Sun, Yan, Ma, Ren, Yin, and Ren}]{sun2023ChatGPT}
Weiwei Sun, Lingyong Yan, Xinyu Ma, Pengjie Ren, Dawei Yin, and Zhaochun Ren. 2023.
\newblock \href {https://arxiv.org/abs/2304.09542} {Is {{ChatGPT Good}} at {{Search}}? {{Investigating Large Language Models}} as {{Re-Ranking Agent}}}.
\newblock \emph{Preprint}, arXiv:2304.09542.

\bibitem[{Thakur et~al.(2021)Thakur, Reimers, R{\"u}ckl{\'e}, Srivastava, and Gurevych}]{thakur2021BEIR}
Nandan Thakur, Nils Reimers, Andreas R{\"u}ckl{\'e}, Abhishek Srivastava, and Iryna Gurevych. 2021.
\newblock \href {https://doi.org/10.48550/arXiv.2104.08663} {{{BEIR}}: {{A Heterogenous Benchmark}} for {{Zero-shot Evaluation}} of {{Information Retrieval Models}}}.

\bibitem[{Touvron et~al.(2023)Touvron, Lavril, Izacard, Martinet, Lachaux, Lacroix, Rozi{\`e}re, Goyal, Hambro, Azhar, Rodriguez, Joulin, Grave, and Lample}]{touvron2023LLaMA}
Hugo Touvron, Thibaut Lavril, Gautier Izacard, Xavier Martinet, Marie-Anne Lachaux, Timothee Lacroix, Baptiste Rozi{\`e}re, Naman Goyal, Eric Hambro, Faisal Azhar, Aurelien Rodriguez, Armand Joulin, Edouard Grave, and Guillaume Lample. 2023.
\newblock {{LLaMA}}: {{Open}} and {{Efficient Foundation Language Models}}.
\newblock \emph{arXiv preprint}.

\bibitem[{Weller et~al.(2024)Weller, Lawrie, and Van~Durme}]{weller2024NevIR}
Orion Weller, Dawn Lawrie, and Benjamin Van~Durme. 2024.
\newblock \href {https://doi.org/10.48550/arXiv.2305.07614} {{{NevIR}}: {{Negation}} in {{Neural Information Retrieval}}}.
\newblock \emph{Preprint}, arXiv:2305.07614.

\bibitem[{Wolf et~al.(2020)Wolf, Debut, Sanh, Chaumond, Delangue, Moi, Cistac, Rault, Louf, Funtowicz, Davison, Shleifer, von Platen, Ma, Jernite, Plu, Xu, Le~Scao, Gugger, Drame, Lhoest, and Rush}]{wolf-etal-2020-transformers}
Thomas Wolf, Lysandre Debut, Victor Sanh, Julien Chaumond, Clement Delangue, Anthony Moi, Pierric Cistac, Tim Rault, Remi Louf, Morgan Funtowicz, Joe Davison, Sam Shleifer, Patrick von Platen, Clara Ma, Yacine Jernite, Julien Plu, Canwen Xu, Teven Le~Scao, Sylvain Gugger, Mariama Drame, Quentin Lhoest, and Alexander Rush. 2020.
\newblock \href {https://doi.org/10.18653/v1/2020.emnlp-demos.6} {Transformers: State-of-the-art natural language processing}.
\newblock In \emph{Proceedings of the 2020 Conference on Empirical Methods in Natural Language Processing: System Demonstrations}, pages 38--45, Online. Association for Computational Linguistics.

\bibitem[{Zhang et~al.(2023)Zhang, Hofst{\"a}tter, Lewis, Tang, and Lin}]{zhang2023RankwithoutGPT}
Xinyu Zhang, Sebastian Hofst{\"a}tter, Patrick Lewis, Raphael Tang, and Jimmy Lin. 2023.
\newblock \href {https://doi.org/10.48550/arXiv.2312.02969} {Rank-without-{{GPT}}: {{Building GPT-Independent Listwise Rerankers}} on {{Open-Source Large Language Models}}}.
\newblock \emph{Preprint}, arXiv:2312.02969.

\bibitem[{Zhu et~al.(2023)Zhu, Yuan, Wang, Liu, Liu, Deng, Dou, and Wen}]{zhu2023Large}
Yutao Zhu, Huaying Yuan, Shuting Wang, Jiongnan Liu, Wenhan Liu, Chenlong Deng, Zhicheng Dou, and Ji-Rong Wen. 2023.
\newblock \href {https://doi.org/10.48550/arXiv.2308.07107} {Large {{Language Models}} for {{Information Retrieval}}: {{A Survey}}}.
\newblock \emph{Preprint}, arXiv:2308.07107.

\bibitem[{Zhuang et~al.(2023{\natexlab{a}})Zhuang, Qin, Hui, Wu, Yan, Wang, and Berdersky}]{zhuang2023Yes}
Honglei Zhuang, Zhen Qin, Kai Hui, Junru Wu, Le~Yan, Xuanhui Wang, and Michael Berdersky. 2023{\natexlab{a}}.
\newblock \href {https://arxiv.org/abs/2310.14122} {Beyond {{Yes}} and {{No}}: {{Improving Zero-Shot LLM Rankers}} via {{Scoring Fine-Grained Relevance Labels}}}.
\newblock \emph{Preprint}, arXiv:2310.14122.

\bibitem[{Zhuang et~al.(2023{\natexlab{b}})Zhuang, Zhuang, Koopman, and Zuccon}]{zhuang2023Setwise}
Shengyao Zhuang, Honglei Zhuang, Bevan Koopman, and Guido Zuccon. 2023{\natexlab{b}}.
\newblock \href {https://doi.org/10.48550/arXiv.2310.09497} {A {{Setwise Approach}} for {{Effective}} and {{Highly Efficient Zero-shot Ranking}} with {{Large Language Models}}}.
\newblock \emph{Preprint}, arXiv:2310.09497.

\end{thebibliography}

\appendix

\section{Additional Usage Examples}
\label{app:example}

\paragraph{Using different LLMs} LLM4Ranking supports easy switching between different LLMs. For example, when using the API model compatible with OpenAI SDK, one can use the following code:
\begin{minted}[fontsize=\small, breaklines]{python}
from llm4ranking import Reranker

reranker = Reranker(
    reranking_approach="rankgpt",
    model_type="openai",
    model_name="gpt-4o", # or other models like "deepseek-v3"
    model_args={"api_key": "sk-xxxxx"}
)
\end{minted}
In contrast, when using the open source model, only a few of arguments need to be changed:
\begin{minted}[fontsize=\small, breaklines]{python}
reranker = Reranker(
    reranking_approach="rankgpt",
    model_type="hf",  # or "vllm"
    model_name="Qwen/Qwen2.5-7B-Instruct",
)
\end{minted}

\paragraph{Customizing Ranking Model}

We provide an example of an ensemble pointwise model to show how to customize the ranking model. As shown in Section~\ref{sec:overview}, the \texttt{PointwiseReranker} takes the argument \texttt{ranking\_func} to pass the model in the \texttt{rerank} function, then we just need to implement a function that satisfies the interface. For example, given the implemented methods Relevance-Generation and Query-Generation, we can use a new class to ensemble them:
\begin{minted}[fontsize=\small, breaklines]{python}
from llm4ranking.model import RelevanceGeneration, QueryGeneration

class EnsemblePointwise:
    def __init__(self, **kwargs):
        self.rg = RelevanceGeneration(**kwargs)
        self.qg = QueryGeneration(**kwargs)

    def __call__(self, query: str, doc: str) -> float:
        score_1 = self.rg(query, doc)
        score_2 = self.qg(query, doc)
        return score_1 + score_2
\end{minted}
Then we can integrate it in the pointwise ranking logit (implemented in \texttt{PointwiseReranker}) and rerank the documents:
\begin{minted}[fontsize=\small, breaklines]{python}
from functools import partial
from llm4ranking.ranker import PointwiseReranker

ranker = PointwiseReranker()
rerank_func = EnsemblePointwise(
    model_type="hf",
    model_name="Qwen/Qwen2.5-7B-Instruct",
)

custom_rerank = partial(
    ranker.rerank,
    ranking_func=ranking_func
)
custom_rerank(
    query: "query text",
    candidates: ["doc0", "doc1", "doc2", ...],
)

>> ["doc2", "doc0", "doc1", ...]
\end{minted}
Benefiting from the flexible implementation of the framework, users can customize their reranking model in a similar way. Similarly, users can customize ranking logic except for the pointwise, such as tourrank or others, and make diversified combinations.

\paragraph{Evaluation} Expect for running evaluation in the command line, it's also possible to use a function:
\begin{minted}[fontsize=\small, breaklines]{python}
from llm4ranking.evaluation.evaluator import simple_evaluate

results = simple_evaluate(
    model_type="hf",
    model_args={"model": "Qwen/Qwen2.5-7B-Instruct"},
    datasets=["dl19"],
    reranking_approach="rankgpt",
    retriever="bm25",
    topk=100,
)
\end{minted}

\section{Experiment Details}
\label{sec:appendix-exp}

\subsection{Evaluation Settings}

In the evaluation experiments, we rerank the top 100 candidates obtained from BM25 and use nDCG@10 as the metric to evaluate the reranking results. For RankGPT, we followed~\citet{sun2023ChatGPT} and set the window size to 20 and step to 10. For TourRank, different from~\citet{chen2024TourRank}, we only performed 1 time of tournament, while more tournaments are expected to further improve the ranking performance.

\subsection{Training Details}

For listwise models, the training data is sourced from~\citet{pradeep2023RankZephyr} which is generated by RankGPT-4 and used for distillation, incuding about 5k samples, and the training process spans three epochs. The learning rate is set to 5e-6, following a cosine decay schedule with a warmup ratio of 3\%.

For pointwise models, the training data is from MS MARCO training set and we used about 24k samples for training. The training setup incorporates three negative samples per instance and we used cross entroy loss to optimize the model. The learning rate is set to 5e-6, following a cosine decay schedule with a warmup ratio of 3\%.

Based on the size of the model, we selected different batch size to fit the memory usage. For all training processes, mixed precision and DeepSpeed is used to optimize memory usage and computational efficiency. All training experiments are conducted on 4 Nvidia A100 GPUs.

\end{document}